\documentclass[runningheads]{llncs}

\usepackage[T1]{fontenc}
\usepackage{graphicx}
\setkeys{Gin}{draft=false}

\usepackage{hyperref}
\usepackage{subcaption}
\usepackage{orcidlink}
\usepackage{color}

\usepackage[svgnames,table]{xcolor}
\usepackage{amsmath}
\usepackage[nameinlink]{cleveref}
\usepackage{utfsym}
\newcommand{\symcheck}{\usym{2713}}

\usepackage{overpic}
\usepackage{booktabs}
\usepackage{multirow}
\usepackage{hyperref}
\usepackage{forloop}

\NewDocumentCommand{\sd}{o}{\mbox{systemd\IfValueT{#1}{-#1}}}

\newcommand{\anonrepo}{https://github.com/crocs-muni/tpmspy}
\newcommand{\rqref}[1]{\hyperlink{rq#1}{\textbf{RQ#1}}}

\usepackage{tcolorbox}
\tcbuselibrary{skins}
\newtcolorbox{rqbox}[1]{enhanced,
	attach boxed title to top center={yshift=-2.2mm},
	fonttitle=\sffamily\bfseries,
	coltitle=black,
	boxed title style={colframe=white,colback=white,top=-0.2em,bottom=-0.4em},
	colback=white,
	sharp corners,
	boxrule=0.3mm,
	title={#1},
	left=2mm,
	right=2mm,
	bottom=2mm,
	enlarge top by=-0.5em,
}

\usetikzlibrary{arrows.meta}

\tikzset{
	pointer/.style = {
		-{Stealth[scale=0.8]}, Crimson, line width=0.8pt,
	},
	attention/.style = {
		Crimson, line width=0.8pt, fill=none, radius=5pt,
	}
}

\begin{document}

\title{TPMSpy: Validation of Measured Boot Systems by Low-Level Tracing of TPM Usage}

\titlerunning{TPMSpy}

\author{
	Roman Lacko\inst{1}\orcidlink{0009-0002-1775-205X}
	\and
	Petr \v{S}venda\inst{1}\orcidlink{0000-0002-9784-7624}
}

\authorrunning{R. Lacko \and P. \v{S}venda}

\institute{Masaryk University, Brno, Czechia}

\maketitle

\begin{abstract}
Measured Boot extends trust in a booted system by recording cryptographic measurements of executed software and system state into a Trusted Platform Module (TPM), enabling subsequent verification through remote attestation.
Although this mechanism is increasingly deployed in contemporary operating systems, its practical security depends on whether implementations measure the expected components under the expected conditions, yet this is not checked systematically.

We propose a platform-agnostic method for analysing low-level TPM usage at the level of virtualized system--TPM interactions.
It enables independent reconstruction and validation of the TPM Event Log without relying on the quoting mechanism itself.
Because it does not depend on implementation details, it is applicable to both open and closed systems.
We demonstrate the method on both Linux and Windows and conduct a systematic longitudinal analysis of Linux systems with \sd{} versions 245--258 (2020--2025), examining how Measured Boot usage evolved and observing wide divergence.
No single usage pattern emerged amongst systems, warranting customized analysis.

The analysis identifies undocumented behavioural changes, reveals inconsistent measurements of user-space \sd{} services, which prevent reliable remote attestation and LUKS disk decryption on such systems. 

\end{abstract}

\section{Introduction}
\label{sec:introduction}

Ensuring the boot integrity of a computing system is one element of an efficient defence against malware~\cite{parno_bootstrapping_2011}.
Firmware and bootloaders can be compromised to render the system inoperable, exfiltrate confidential information, or gain unrestricted access~\cite{ruan_boot_2014}.
Two general approaches are used to address this: \textit{Verified Boot} ensures only known components are allowed to run, while \textit{Measured Boot} collects evidence in the form of the identities of executed components~\cite{ling_secure_2021}.
The collected evidence is then evaluated during remote attestation~\cite{coker_principles_2011}.
Especially on personal computers, the evidence is often used to unlock encrypted volumes automatically---BitLocker on Windows~\cite{microsoft:bitlocker} and LUKS on Linux~\cite{docs:luks}.
While Measured and Verified Boot, in principle, promise an effective defence against early-boot threats, their implementations evolve, and the behaviour of these mechanisms differs across vendors, versions, and configurations, necessitating a non-intrusive, scalable way to verify their properties as they evolve.

Common implementations of Verified and Measured Boot in PCs and servers utilize a Trusted Platform Module (TPM)~\cite{tcg:tpm}, a secure coprocessor with tamper-resistant key storage and dedicated measurement registers.
Every boot component participating in Measured Boot measures the component that will be executed next in the chain.
The measurements are passed to TPM, which incorporates them into the collected evidence using cryptographic hash functions, preventing a malicious component from erasing its own traces retrospectively.
The appraisers request a copy of the registers signed by the TPM, called a TPM Quote.
The system maintains a TPM Event Log that includes inputs and metadata for each measurement, which is intended to aid in interpreting the TPM Quote.
While the correctness of the log can be verified by replaying the events and comparing the results with the evidence, a scalable and reliable method for verifying completeness (i.e., that all components were measured with matching events recorded in the log) is not yet available.
Most boot components lack a precise description of their TPM usage, and directly observing TPM interactions to reconstruct independent evidence is technically challenging.
Our paper aims to identify methods that allow this independent reconstruction.

The \sd{} project~\cite{github:systemd} has been documenting its TPM utilization since version 248.
Unlike other systems, which rarely disclose such details (notably Windows~\cite{microsoft:pcr-banks}), this allows for transparent development of attestation policies.
Still, the documentation must match the behaviour, which aligns with the goal of this paper: verifying the completeness of the measurements.

We investigate the following research questions:

\begin{rqbox}{Research Questions}
	\hypertarget{rq1}{\textbf{RQ1:}} Can we independently reconstruct and validate the TPM Event Log sent with the signed TPM Quote without relying on the quoting mechanism?

	\hypertarget{rq2}{\textbf{RQ2:}} Can we automatically assess how the Linux ecosystem built on \sd{} utilizes Measured Boot, and how it has evolved?

	\hypertarget{rq3}{\textbf{RQ3:}} Do the uses of TPM Platform Configuration Registers (PCR) captured in RQ2 correspond to the documented behaviour of \sd{}?
\end{rqbox}

\noindent \textbf{Contributions.} To answer our research questions, we provide the following:

\begin{enumerate}
	\item Survey of principal options for TPM usage tracing (\Cref{sec:options-tracing}).
	\item Platform-agnostic and automatic method based on a virtualized system with a TPM, allowing to capture reads and writes between measured components and the TPM, and thus enabling the reconstruction and validation of the TPM Event Log (\Cref{sec:technical-realization}).
	\item Systematic analysis of TPM access patterns of all major versions of \sd{}-based Linux distributions between 2020 (v245, before it had started to utilize TPM) and 2025 (v258)
	      (Sections \ref{sec:methodology} and \ref{subsec:pcr_evolution}).
	\item Detection and analysis of an inconsistency in the expected behaviour in TPM Event Log when measuring encrypted root volume, effectively rendering remote attestation ineffective for such setups (\Cref{subsec:setup_uki}).
\end{enumerate}

All source code and experiment configurations are released under an open-source license at \anonrepo{}.
\section{Background: Protecting the Integrity of Boot Process}
\label{sec:background}

The firmware (BIOS or UEFI~\cite{uefi_forum_unified_2024}) initializes the hardware and bootstraps the operating system~\cite{ruan_boot_2014}.
The firmware can detect malfunctioning hardware and firmware, but it is not designed to detect malicious components, such as rootkits.
\textit{Verified Boot} and \textit{Measured Boot}~\cite{dietrich_secure_2008,ruan_boot_2014,smith_boot_2024} are general mechanisms built on top of a secure hardware component to protect against these issues.

\subsection{Verified and Measured Boot}

Both methods rely on one or more Roots of Trust (RoT), trusted components (usually hardware) within the computer, with specific roles.
The key difference is that while Verified Boot enforces a priori verification based on certified components, Measured Boot collects evidence and enables a posteriori verification by a remote party.

\textbf{Verified Boot} or \textit{Secure Boot} first executes the RoT, which validates signatures of firmware and its configuration against embedded keys or certificates~\cite{dietrich_secure_2008,khalid_implementing_2013,profentzas_performance_2019,ruan_boot_2014}.
A violation of this procedure usually halts the boot process and requires the administrator to investigate.

In \textbf{Measured Boot} or \textit{Trusted Boot}, multiple RoTs performing different roles exist.
The RoT for Measurement starts by measuring itself, usually by hashing its own code~\cite{berger_scalable_2015,chevalier_bootkeeper_2019,huang_research_2016,smith_boot_2024,yeluri_platform_2014}.
The measurements are forwarded to the RoT for Reporting, which accumulates them using a cryptographic operation ``extend'' rather than storing them directly.
The entire system is then measured iteratively: the currently running component must measure the next component before passing control~\cite{mitchell_trusted_2005,smith_trusted_2005}.
This constructs a chain of measurements, also a chain of trust, that can be later inspected by a remote party during a remote attestation~\cite{coker_principles_2011} by comparing the expected chain of hashes for the expected components.
Based on the reported values, the party decides whether the system has reached a reliable state to perform sensitive tasks or is instead rejected and must be inspected.

In principle, Measured Boot does not prevent malicious firmware and hardware from gaining control.
However, if implemented correctly, the measurements of malicious components are also included in the collected evidence and cannot be retrospectively altered or erased without breaking the cryptographic properties imposed by the RoT.

In practice, combinations of both methods can be deployed, such as \textit{Trusted Boot} in Microsoft Windows~\cite{microsoft:secure-boot}, which builds on Verified Boot, but uses elements of Measured Boot during system initialization.

\subsection{Measured Boot Based on Trusted Platform Module}
\label{subsec:bg_mb_tpm}

The Trusted Platform Module (TPM) is a specification for trusted hardware designed by the Trusted Computing Group and currently at version 2.0. While it can serve as a generic secure coprocessor~\cite{tcg:tpm}, this section focuses on aspects that the TPM provides for implementing Measured Boot.
The secure component implicitly trusted to measure itself reliably is called Core RoT for Measurement (CRTM), which passes these measurements to TPM.
In turn, the TPM fulfils the role of RoT for Reporting, as it features 24 Platform Configuration Registers (PCRs) designed to securely collect and report the measurements.
Components cannot write to the registers directly; instead, TPM exposes a \textit{PCR Extend} operation, where a hash of the register's previous value and the new measurement is stored in the register, i.e. $\text{PCR}[n] \leftarrow h(\text{PCR}[n] \mid\mid \text{data})$.
A single PCR is typically extended multiple times during a boot, accumulating contributions from several components assigned to the same register (\Cref{fig:pcr-tpm-spy}).
When the measurements are finalized, a third party can request the values from PCR banks signed by the TPM's private key, called a TPM Quote.
The system maintains the TPM Event Log, which contains metadata for each measurement intended to aid interpretation of the TPM Quote.

In theory, during remote attestation, the appraiser links the evidence~\cite{coker_principles_2011,tcg:guidance}:

\begin{enumerate}
	\item The Quote signed by a TPM, its certificate anchor is a trusted manufacturer.
	\item The TPM Event Log records match the hash values of known components.
	\item Simulation of the TPM Event Log yields the values in the TPM Quote.
\end{enumerate}

Unless the TPM's private key is known to a malicious party, it should not be possible to craft a valid signature for a fabricated TPM Quote.
The TPM Event Log then supports the collected evidence by explaining how the values were computed.
Due to TPM's memory limitations, the TPM Event Log is maintained by the system and must be handled carefully when passing execution context from the firmware to the operating system.
\section{Methods for Analysing Measured Boot}
\label{sec:options-tracing}

To verify the validity of the TPM Event Log (our \rqref{1}), it is necessary to provide independent evidence of the measurements recorded during system boot.
The ideal method of capturing such evidence does not interfere with or interact with the processes running in the system, but can observe or validate all the interactions with the TPM.

\subsection{Criteria for Method Selection}
\label{subsec:mb_criteria}
Measured Boot relies on Roots of Trust to record the measurements and hand them over when requested.
The state reached by this process must follow from the TPM's initial state and the measurements it received.
Based on this observation, we devised four main criteria for selecting an appropriate method.

The ideal method must observe the components performing the measurements, ideally during every stage of the boot process (\textit{Coverage}).
Once set up, the method should enable repeatable executions and facilitate quick validation of the measurements (\textit{Deployability}).
This targets practical use in development and Continuous Integration environments, where developers can validate that their changes align with expected behaviour.

For the final properties, the method must not require modification of the observed system (\textit{Intrusive}) or rely on its proper function (\textit{Independent}).
Such dependencies could compromise the method's integrity when malicious components are present.

We propose a method that observes the system in a virtualized environment, where communication with RoT is limited to a channel we can intercept and analyse.
Platforms implementing Measured Boot on top of a Trusted Platform Module (TPM) fulfil our requirements.
The properties and limitations of these approaches are discussed below, with summarization in \Cref{table:summary_of_methods}.

\subsection{Evaluated Methods for Measured Boot Analysis}

In this section, we discuss five principal base approaches that can be utilized to observe the measurements: 1) TPM Event Log analysis, 2) formal verification of code, 3) hardware probes, 4) software probes, and 5) virtualization.

\subsubsection{TPM Event Log analysis} is a baseline method for assessing the validity of measurements collected during the boot process and is already established as the core of remote attestation with TPM~\cite{coker_principles_2011,tcg:guidance}.
An appraiser obtains the TPM Quote, signed with the TPM's private key, from the target system, along with the TPM Event Log.
By simulating the events recorded in the log and comparing the results with the quote, the appraiser can verify the authenticity of the measurements.

This approach is non-invasive and well established in practice, but does not address \rqref{1}, as it does not provide evidence independent of the platform.
It has been demonstrated that a malicious party can reset certain types of TPM without affecting the platform and re-measure parts of the system, thereby evading detection~\cite{han_bad_2018,winter_hijackers_2012}.
Furthermore, the log format accounts for constraints during early boot stages (such as memory and storage yet to be initialized) and limits the amount of metadata that can be stored.
Any method that relies solely on the TPM Event Log inherits these constraints; therefore, an independent observation outside the mechanism is required.

\subsubsection{Formal verification} can describe Measured Boot as a set of logical statements that can be used to validate the behaviour of the components participating in the boot process~\cite{grimm_survey_2018}.
The approach works well for verifying runtime or security properties of components with explicitly and rigorously specified tasks, such as firmware and its interactions~\cite{ray_formal_2019}, CRTM~\cite{tao_dice_2021}, or bootloaders~\cite{gordon_applying_2025}.
Furthermore, efforts are underway to extend verification to more complex software, including kernel drivers~\cite{chen_veld_2024} and kernels themselves~\cite{chen_atmosphere_2023,klein_sel4_2009,de_oliveira_efficient_2019}.
On the other hand, while methods for the formal analysis of every link in the boot chain exist~\cite{vasudevan_formal_2024,yuan_verified_2021}, they are primarily limited to embedded systems, where the boot chain is comparatively smaller than that of PCs or servers.
Analysis in more general settings is further complicated by the widespread use of hardware with closed-source firmware and by user-space services' ability to participate in the boot process.

Therefore, formal verification is theoretically sufficient for answering \rqref{1}, but applying it to the full boot chain of a general-purpose PC remains an unsolved problem.

\subsubsection{Hardware probes} are devices attached to an exposed data bus or power network that can observe or alter signals.
This allows for analysing components' behaviour directly (including TPM requests)~\cite{kursawe_analyzing_2025} or revealing secret data or performing unintended operations~\cite{anderson_tamper_1996}.
Vulnerabilities in TPMs as physical, discrete chips (dTPMs) have been discovered by employing this method: an attacker can force-reset the TPM and reconstruct fabricated measurements that hide their presence~\cite{han_bad_2018,winter_hijackers_2012,winter_hijackers_2013} or bypass its security measures~\cite{yli-mayry_automated_2024}.
These attacks are not limited to dTPMs; TPMs running as firmware in the CPU's trusted execution environment (fTPMs) can be attacked by manipulating the power network~\cite{jacob_faultpm_2023}.
Furthermore, eavesdropping on dTPM communication packets on the data bus has been demonstrated~\cite{kursawe_analyzing_2025}.

Inexpensive, open-source probes designed for dTPMs already exist (e.g., TPM Genie~\cite{github:tpmgenie}), allowing this method to be considered for dTPM chips at the very least.
However, the method requires physical access to the system to attach such a probe to a bus.
In the case of fTPMs, direct interaction with the system is required to inject faults and expose the TPM's internal state.
Finally, recent TPMs integrated into CPUs' circuit boards (iTPMs), such as Microsoft Pluton~\cite{microsoft:pluton}, have a bus that is inaccessible without dismantling the chip casing, which is already a highly invasive and delicate process.

\subsubsection{Software probes} can intercept the data from measurement software before it is passed on to the TPM communication bus.
Linux Kernel already provides several interfaces for tracing, notably function entry and exit probes~\cite{docs:fprobe}, more general kernel instrumentation probes~\cite{docs:kprobes}, and Berkeley Packet Filter subsystem~\cite{docs:bpf}.
While they are mainly designed for behaviour analysis, they can be utilized to detect malware~\cite{caviglione_kernel-level_2021,vurdelja_detection_2020}.
Similar tracing mechanisms exist for Microsoft Windows~\cite{lanzi_k-tracer_2009,shan_tracer_2011}, but are limited in scope since its source code is not disclosed.
We will narrow this discussion to the Linux kernel, which provides an alternative option: operating systems expose abstract, high-level interfaces for communicating with devices via drivers.
Their source code can be modified to intercept the measurements.

The applicability of this approach breaks down with the components that execute before the kernel, as there is no such unifying interface.
Additionally, the closed-source firmware prevents capturing measurements from the early stages of the boot.
Even open-source early-boot firmware cannot be traced by kernel tracers and must be modified, which increases the complexity and volatility of this approach.
Furthermore, such a method must rely on the system being rootkit-free, as malware running with elevated privileges could interfere with measurement capture.
Ultimately, this method is feasible for measurements submitted by the kernel, but it falls short of reliably capturing early-boot measurements.

\subsubsection{Virtualization} of both the system and TPM.
The methods discussed so far are powerful but challenging to scale, automate, and interpret while minimizing interference with the observed system.
Virtual TPMs (vTPMs) have been extensively researched since the advent of cloud computing, with the goal of providing a TPM instance for each guest, backed by real secure hardware~\cite{de_benedictis_novel_2024,sadeghi_property-based_2008,shi_security-improved_2015}, or of developing and testing TPM software~\cite {haas_state_2024}.
In such a setup, the hypervisor mediates all communication between the virtual machine and the TPM backend, which is then fully observable from outside of the guest.
Instead of a hardware TPM, the setup can be built on top of a software TPM (swTPM)~\cite{github:swtpm,strasser_software-based_2008}.
This allows for independent observation of all TPM interactions, fulfilling the key requirement of \rqref{1}, without altering the hardware or modifying the software.
And since the measurements originate from the system rather than the TPM, the implementation differences between swTPM and real TPMs are out of scope for \rqref{1}.
This approach overall provides a better balance between independence and practicality than previous methods.

Prior methods enable either the validation of specific components or the observation of TPM measurements as summarized in \Cref{table:summary_of_methods}.
Neither provides an independent, non-intrusive and scalable validation of the complete TPM Event Log.
Virtualization provides most of the desired properties, except for the observation of the TPM behaviour, which we aim to fill in with this research.

\newcommand{\tbnay}{No}
\newcommand{\tbyay}{Yes}

\begin{table}
    \centering
    \begin{tabular}{l c c c c}
        Method                   & Independent & Intrusive & Coverage            & Deployability \\
        \midrule
        TPM Event Log Analysis   & \tbnay       & \tbnay        & Full                & High \\
        Formal verification      & \tbyay       & \tbnay        & Isolated components & High \\
        Hardware probes          & \tbyay       & \tbyay        & Full                & Low  \\
        Software probes          & \tbnay       & \tbyay        & Partial             & Low  \\
        Virtualization           & \tbyay       & \tbnay        & Full                & High \\
    \end{tabular}
    \vspace{1em}
    \caption{Summary of methods for analysing Measured Boot based on TPM. The properties (columns) are discussed in \Cref{subsec:mb_criteria}.}
    \label{table:summary_of_methods}
\end{table}

\subsection{Technical Realization of TPM Measurement Tracing}
\label{sec:technical-realization}

This section describes a practical implementation of the independent TPM Event Log validation to answer \rqref{1}, based on the previous discussion.
The system consists of a virtualized platform and a separate software-emulated TPM.
All TPM commands and responses in the guest are translated into messages passed between the hypervisor and TPM emulator processes.
As discussed in the previous section, this approach already makes the validation \textit{non-intrusively} \textit{independent} of the configuration and software inside the virtual machine, guarantees \textit{complete} observation of the boot process measurements, and can be \textit{deployed} repeatably for any virtualized system.
However, no existing tool applies this to validate the TPM Event Log.

The interaction between a virtualized system and a software-emulated TPM (swTPM) utilizes the following procedural steps:

\begin{enumerate}
    \item swTPM exposes a communication channel, e.g. a UNIX socket.
    \item When a guest system is started, the hypervisor connects to this channel.
    \item TPM commands from the guest then appear on this channel as data flow.
\end{enumerate}

\begin{figure}
	\centering
	\includegraphics[width=\textwidth]{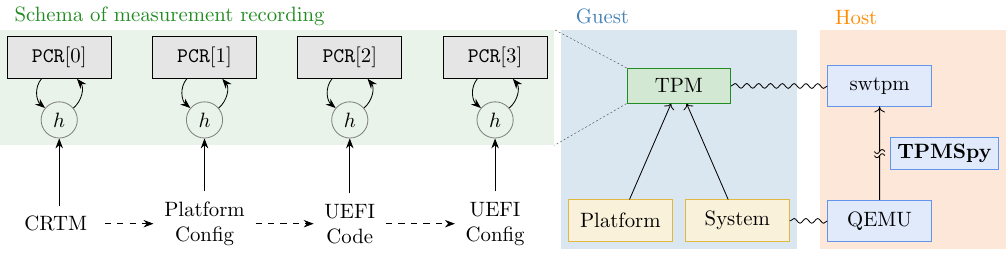}
	\caption{
		Schema of Measured Boot implemented with a software TPM.
        PCR banks (4 of 24 shown) reside inside the TPM and are extended only by hashing the input and the bank's previous value ($h$).
        The operation starts with CRTM, which measures its configuration, passes control to UEFI, etc.
        In our setup, the Guest system is emulated by QEMU in the Host, except the TPM, which is emulated by swtpm.
        The PCR Extend command (represented by an upward arrow in Guest) is translated into data sent over a socket (represented by an upward arrow in Host) and intercepted by our TPMSpy component.
	}
	\label{fig:pcr-tpm-spy}
\end{figure}

The second step enables interception by inserting a software interposer in place of the original swTPM channel, acting as a relay that can store, modify, or respond to passing messages (\Cref{fig:pcr-tpm-spy}).

We implemented TPMSpy\footnote{\anonrepo{}} as a concrete implementation of this mechanism.
As depicted in \Cref{fig:pcr-tpm-spy}, it sits on a channel between a guest running in QEMU~\cite{docs:qemu} and a TPM emulator called swtpm\footnote{Here, swtpm is a concrete implementation of a software TPM (swTPM).}~\cite{github:swtpm}.
In practice, the implementation is non-trivial: communication between QEMU and swTPM uses a two-stage protocol where a control channel dynamically negotiates a separate data channel for actual TPM commands, making a fixed-path relay insufficient.
We considered patching swtpm or dynamically overriding a QEMU dispatcher function.
However, both designs depend on the internal implementation of swtpm or QEMU and can break after updates.
Therefore, we ultimately decided to create a helper program that relies on external behaviour: it impersonates swtpm and executes when QEMU sets up a new virtual machine.
Then it starts the real swtpm daemon and TPMSpy, and sets up UNIX sockets so that TPMSpy's socket is what QEMU actually connects to.

TPMSpy features a modular design in which the core handles connections and passes intercepted messages to plug-in modules.
Message handling in the core is not trivial for two reasons.
First, QEMU dynamically creates the TPM data channel and sends it over the control channel to the TPM daemon as an open file descriptor; a UNIX mechanism which allows processes to share resources.
Second, QEMU expects the TPM daemon to handle new sessions as additional VMs are started.
TPMSpy therefore manages one control and one data channel on each side per VM and must correctly label them for later correlation.

Plug-in modules enable future extensions, such as capturing context within the VM or altering measurements to verify the sensitivity of remote attestation tools.
Currently, the module wraps the captured messages with metadata (timestamps, channel identities, and ancillary messages to link related channels for analysis) and stores them in a file.
Additionally, we developed scripts that connect to the VM via SSH and collect metadata relevant to answering \rqref{1} and \rqref{2}: the TPM Event Log, and versions of the bootloader, kernel, and \sd{}.
Another set of scripts facilitates analysis of collected data by generating graphs of PCR modifications over time and variance across multiple runs, and by comparing the TPM Event Log with the TPMSpy trace, flagging discrepancies.

\section{Evaluation Methodology of \sd{} Measurements}
\label{sec:methodology}

This section describes a practical application of the method described in \Cref{sec:technical-realization} to a Measured Boot performed by a widely-used \sd{}-based Linux OS.
The goal is to describe a concrete system for which it is possible to formulate expectations about its use of TPM for the Measured Boot, and changes to this system that would result in observable changes in this behaviour.
The next section will then compare and discuss the expected behaviour with the actual observations made by TPMSpy.

Trusted Computing Group (TCG) reserves PCR 0--7 for the platform and PCR 8--15 for the operating system and services~\cite{tcg:tpm}.
While we captured access to all PCRs, we primarily focused on analysing the behaviour of software that extends PCRs 8--15, based on the documented and inspectable behaviour of firmware and \sd{}.

\subsection{Threat Model}
\label{subsec:threat-model}

We assume the host, the hypervisor, and swTPM are benign, as TPMSpy only aims to observe and analyse the communication between the virtualized system and swTPM.
The method requires the hypervisor to reliably forward guest measurements to the swTPM, and for the latter to correctly process them.
The observed guest system may be compromised, running malicious or undocumented components, or attempting to tamper with its measurements.
Once the guest's TPM Event Log is obtained, TPMSpy can detect discrepancies, such as dropped, added, reordered or modified events, between the TPM Event Log and the observed measurements, including those caused by TPM reset or re-measurement attacks.
However, it cannot detect components that were never measured by the guest system, as such components do not produce an observable event.

\subsection{Analysis of Firmware Behaviour}
\label{subsec:fw}

On a platform that follows the TCG's recommendations, updating the firmware or modifying its configuration should result in different hash values being extended into PCRs 0 or 1.
Similarly, a modification of the bootloader should result in a different value being extended to PCR 4.

Hardware vendors typically use their own proprietary, closed-source firmware with no documentation on TPM use in Measured Boot.
Virtualization hypervisors generally default to open-source firmware (e.g. QEMU uses SeaBIOS for legacy and TianoCore for UEFI boot~\cite{docs:qemu}), but we are not aware of a precise documentation of their PCR utilization being provided either.

This limits an experiment based on firmware modifications: we can only formulate the expected behaviour by observing at least one different value in PCR banks when comparing two different firmware versions.
Beyond this and TCG's recommended description, no further expectations can be stated without auditing the source code of each version of the considered firmware.
For this reason, we capture PCR 0--7 and compare them with other runs to detect unexpected changes, but we leave the interpretation of these measurements outside the evaluation.

\subsection{Analysis of \sd{}'s Documented Behaviour}
\label{subsec:doc}

In a Linux-based operating system, after the kernel is loaded, the first process, called \texttt{init}, is started.
It is tasked with setting up services and the user environment.
The \sd{} project is a well-known \texttt{init} daemon implementation featured in most major contemporary distributions, such as Ubuntu, Debian and Fedora.
Since v248 was released in 2021, the \sd[cryptenroll]'s documentation has maintained a table describing the use of PCR banks by the project's components.
The documentation also mentions other projects that extend PCRs, as it is intended to help system administrators choose PCRs to bind the volume decryption keys to.

Compared to firmware behaviour, this allows us to formulate more detailed expectations; specifically, each version of \sd{} describes the set of PCRs it extends.
To the best of our knowledge, no other Linux init daemon features Measured Boot based on a TPM.

We reviewed major versions of \sd{}'s manual pages in \texttt{man}~\cite{github:systemd} and noted the changes that mention TPM or PCR in \Cref{table:pcrdocs}.
All \sd{} services and user-space tools, except \sd[boot] mentioned in \Cref{table:pcrdocs}, require a specific setup: the kernel image, parameters, and initial RAM file system must be bundled into a Unified Kernel Image (UKI) with \sd[stub].

\newcommand{\tbcheck}{\symcheck}
\newcommand{\lightrule}{\arrayrulecolor{black!25}\specialrule{0.3pt}{0.5pt}{0.5pt}}

\begin{table}
    \centering
    \begin{tabular}{l r c >{\footnotesize}l}
        \sd{}                    & PCR& Default  & Component \\
        \midrule
        v247                     &    &          & \textit{no documented use of TPM up to this version}\\
        \lightrule
        v248                     & 8  & \tbcheck & \sd[boot]\\
        \lightrule
        v249                     &    &          & \textit{no changes}\\
        \lightrule
        \multirow[t]{2}{*}{v250} & 10 &          & Linux Kernel's Integrity Measurement Architecture\textdagger{} (IMA)\\
                                 & 14 &          & Shim\textdagger{}, a trivial UEFI loader\\
        \lightrule
        \multirow[t]{4}{*}{v251} &  4 &          & \sd[stub] if using system extension images\\
                                 &  8 &          & GRUB\textdagger{}; \sd[boot] stops extending this PCR\\
                                 &  9 & \tbcheck & Linux Kernel\textdagger{} initial RAM filesystem since 5.17\\
                                 & 12 & \tbcheck & \sd[boot]; to avoid conflict with GRUB~\cite{systemd:v251-released}\\
        \lightrule
        \multirow[t]{2}{*}{v252} & 11 &          & \sd[stub,pcrphase] if using Unified Kernel Image (UKI)\\
                                 & 13 &          & \sd[sysext] if using extension images\\
        \lightrule
        v253                     & 15 &          & \sd[cryptsetup] if configured for encrypted volumes\\
        \lightrule
        v254                     & 15 &          & \sd[pcrmachine,pcrfs@] if using UKI\\
        \dots                    &    &          & \\
        v258                     &    &          & \textit{no changes since v254}
    \end{tabular}
    \vspace{1em}
    \caption{
        Measurements declared by \sd[cryptenroll] documentation.
        The \textit{Default} column indicates measurements that require no additional configuration when \sd[boot] is the only bootloader.
        Components marked \textdagger{} are not part of \sd{}, but are included in the documentation.
    }
    \label{table:pcrdocs}
\end{table}

\subsection{Experimental Setups}

Based on the above description, we prepared two experimental setups: one to capture the evolution of PCRs extended by each version of \sd{}, and the other to test optional features based on UKI and an encrypted root volume.

For reproducible compilation of older versions of \sd{}, we chose NixOS as the Linux distribution.
One disadvantage of NixOS is that the UKI support required by the user-space tools has only been available in NixOS since January 2024, over a year after \sd{} v252 was released in October 2022.
It supports \sd[cryptsetup], but is still considered experimental and lacks some \sd{} measurement services, such as \sd[pcrmachine].
Therefore we additionally set up and analyzed measurements from Ubuntu 24 LTS (v255), Fedora 41 (v256), 42 (v257) and 43 (v258).

In summary of the expected behaviour:

\begin{enumerate}
	\item \textbf{Base Setup} consists of NixOS with configurations locked to specific versions (called \textit{flakes}) describing versions of \sd{} from v245 to v258.
        Only PCRs 8 and 9 are to be extended up to v250, and 9 and 12 since v251.
        Fedora and Ubuntu are installed with disk encryption enabled, but no specific PCR measurement is set up.
        The behaviour of \sd{} utilities should not differ from NixOS.

	\item \textbf{Encrypted Root Setup} adds Unified Kernel Image (UKI), which is a feature required by \sd[stub] and user-space services to perform additional measurements.
        This setup also takes advantage of the encrypted root, where UKI and a kernel parameter are required for \sd[cryptsetup] to measure the volume encryption key.
	      We expect to see PCRs 11 and 15, in addition to PCRs 9 and 12, extended from the base setup in all experiments.
	      The documentation for v252 also suggests that we should expect more than one measurement in PCR 11 in NixOS.
\end{enumerate}

\begin{figure}
	\centering
	\includegraphics[width=\textwidth]{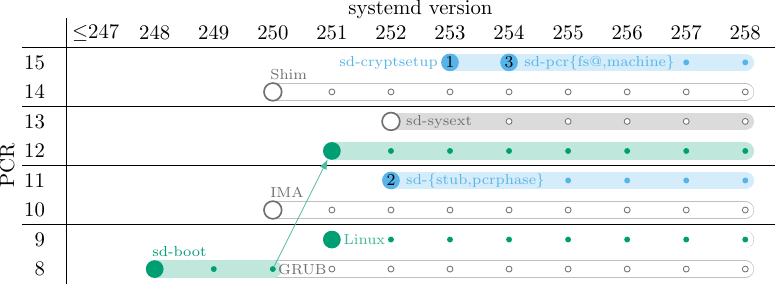}
	\caption{
		Use of PCRs as documented by \sd[cryptenroll].
		Green nodes represent measurements expected in all the experiments.
		Blue nodes are expected only in the experiments with UKI and disk encryption.
		Light nodes represent feature-dependent measurements that are disabled by default.
		A number in the node denotes that there are multiple tools utilizing the PCR.
		Unfilled paths represent measurements by projects that are not part of the \sd{} itself.
	}
	\label{fig:sdusage}
\end{figure}

\Cref{fig:sdusage} displays the evolution of PCR usage as documented by \sd{}.
It also highlights measurements expected in the experiments.

\subsection{Discussion of Expected Behaviour}

Experimental setups are designed to enable automatic capture of multiple system images.
In addition to verifying documented PCR usage, comparing captures of the same version can verify behaviour relevant for remote attestation:

\begin{enumerate}
	\item The list of PCR extend commands in TPM Event Log must match the list in TPMSpy capture.
	\item All captures of the same version must result in one set of PCR values.
	\item The order of operations modifying PCRs must not differ.
\end{enumerate}

To illustrate the last point, if two commands extending the same PCR with different values are reordered, the resulting PCR will have a different hash and can be detected by comparing PCR banks.
But if two consecutive commands targeting different PCRs get swapped in another capture, this will not affect PCR bank values.
Comparing TPM Event Logs will reliably uncover this swap only if each boot component treats the PCR extension command and its TPM Event Log entry as atomic operations, i.e. both are guaranteed to finish before another measurement is started.
TPMSpy does not rely on this requirement, as it intercepts the commands between the system and TPM, and comparing the captures will reliably uncover such swaps.

\section{Experimental Validation of \sd{} Interactions}
\label{sec:evaluation}

We collected measurements from 1,000 executions (boots) of the experimental setup using TPMSpy and \sd{} for each version, ranging from v245 to v258.
We observed an agreement between the captured traces and the TPM Event Logs recovered from the systems.
When \sd{}'s user-space tools (specifically \sd[cryptsetup]) and Integrity Measurement Architecture (IMA) were enabled, we observed that their measurements were not recorded in the same TPM Event Log used by the platform, but in two distinct files.

Comparing the captures with the documentation of \sd{}'s utilities, we identified undocumented measurements in \sd[boot] prior to v248.

\subsection{Setup 1: Evolution of PCR use by \sd{}}
\label{subsec:pcr_evolution}

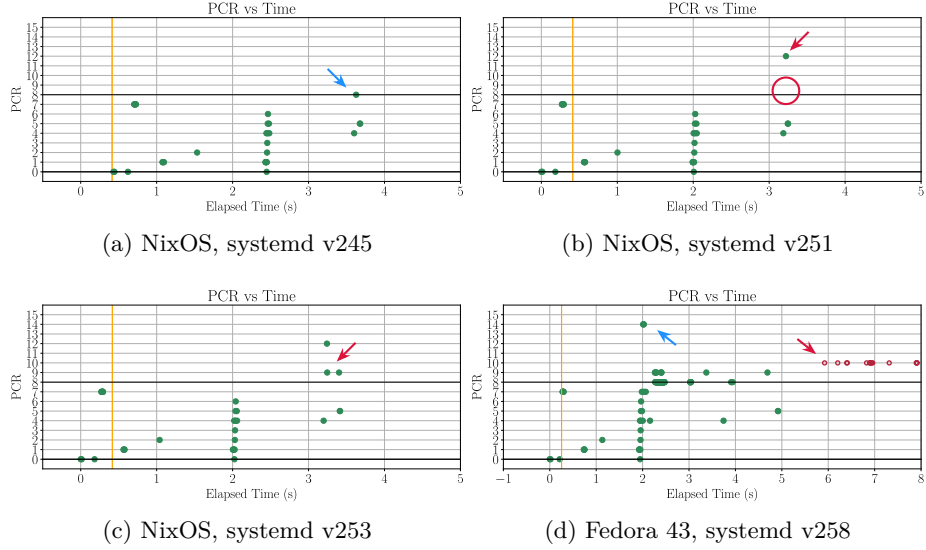
\begin{figure}
	\centering
	\begin{subfigure}[t]{.49\textwidth}
		\centering
		\begin{overpic}[width=1\linewidth,percent]{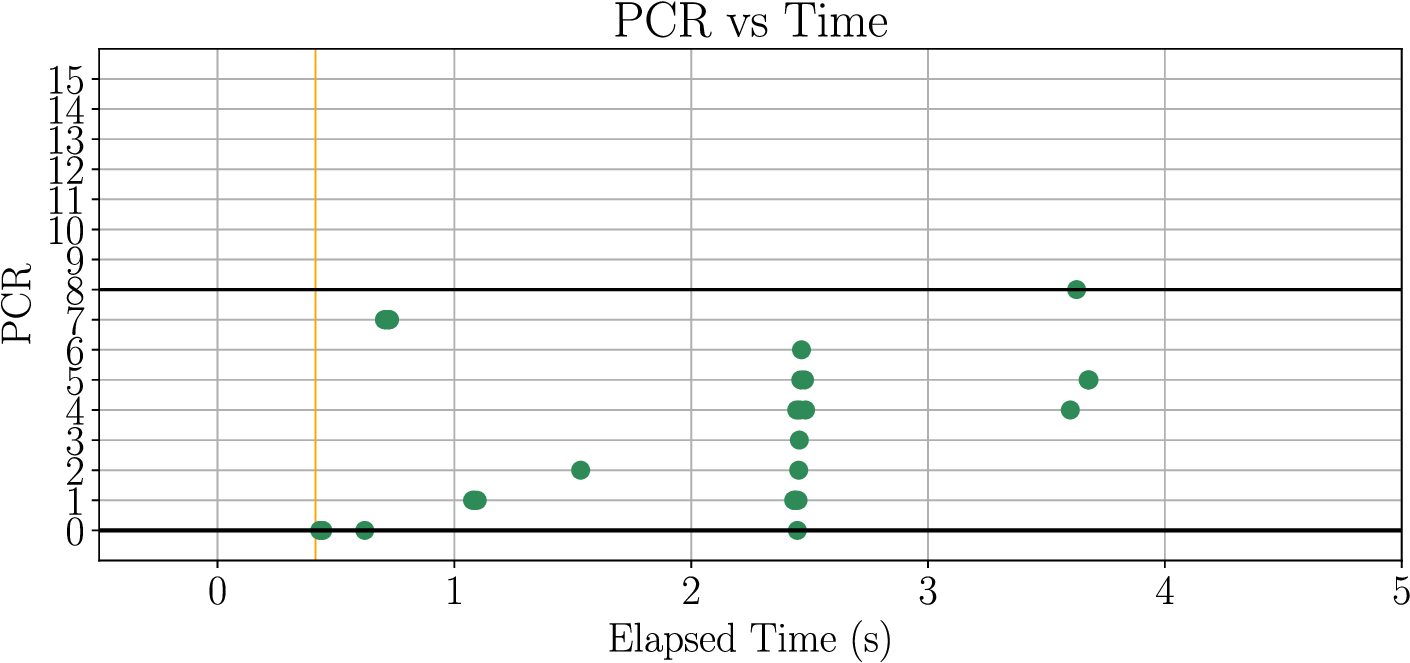}
			\put(64, 28) {
				\begin{tikzpicture}[overlay]
					\draw[pointer,DodgerBlue] (0.25, 0.25) -- (0.5, 0.0);
				\end{tikzpicture}
			}
		\end{overpic}
		\caption{NixOS, \sd{} v245}
        \label{fig:sd245}
	\end{subfigure}
	\begin{subfigure}[t]{.49\textwidth}
		\centering
		\begin{overpic}[width=1\linewidth,grid=false]{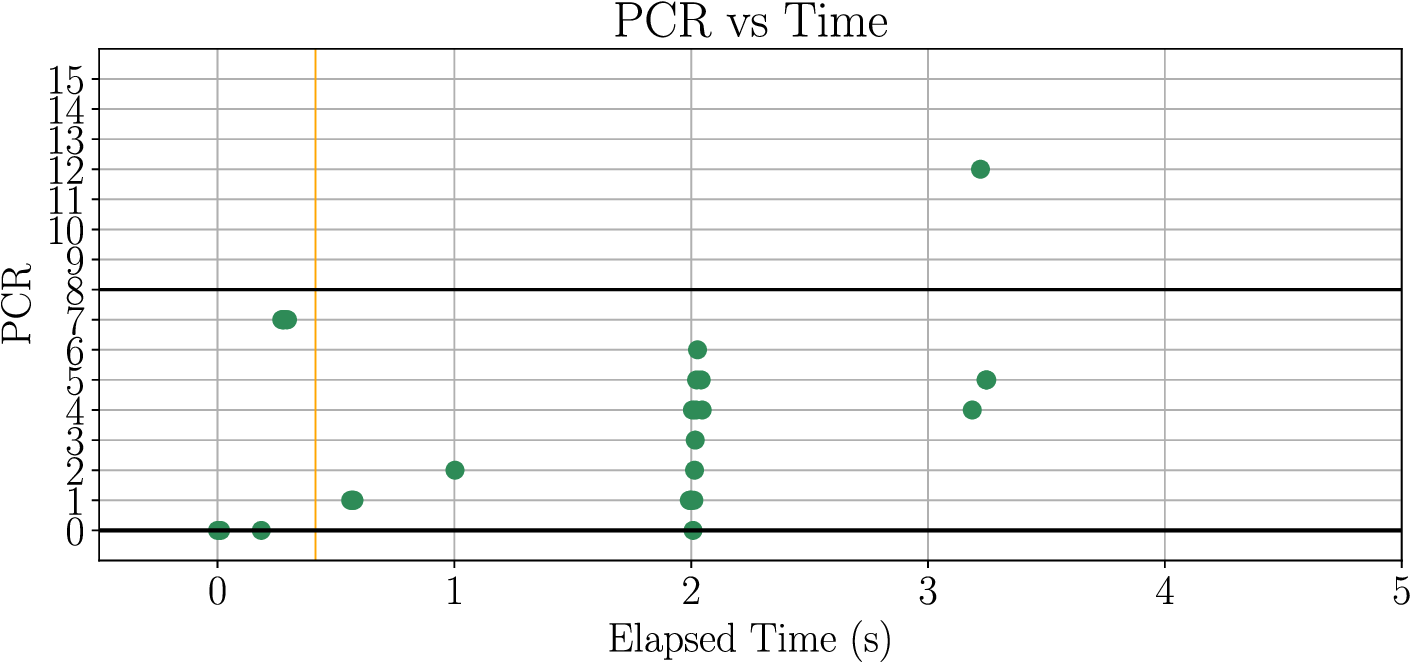}
			\put(59.3, 27) {
				\begin{tikzpicture}[overlay]
					\draw[attention] (0.5, 0.02) circle;
					\draw[pointer] (0.80, 0.80) -- (0.55, 0.55);
				\end{tikzpicture}
			}
		\end{overpic}
		\caption{NixOS, \sd{} v251}
		\label{fig:sd251}
	\end{subfigure}
    \par\bigskip
	\begin{subfigure}[t]{.49\textwidth}
		\centering
		\begin{overpic}[width=1\linewidth,grid=false]{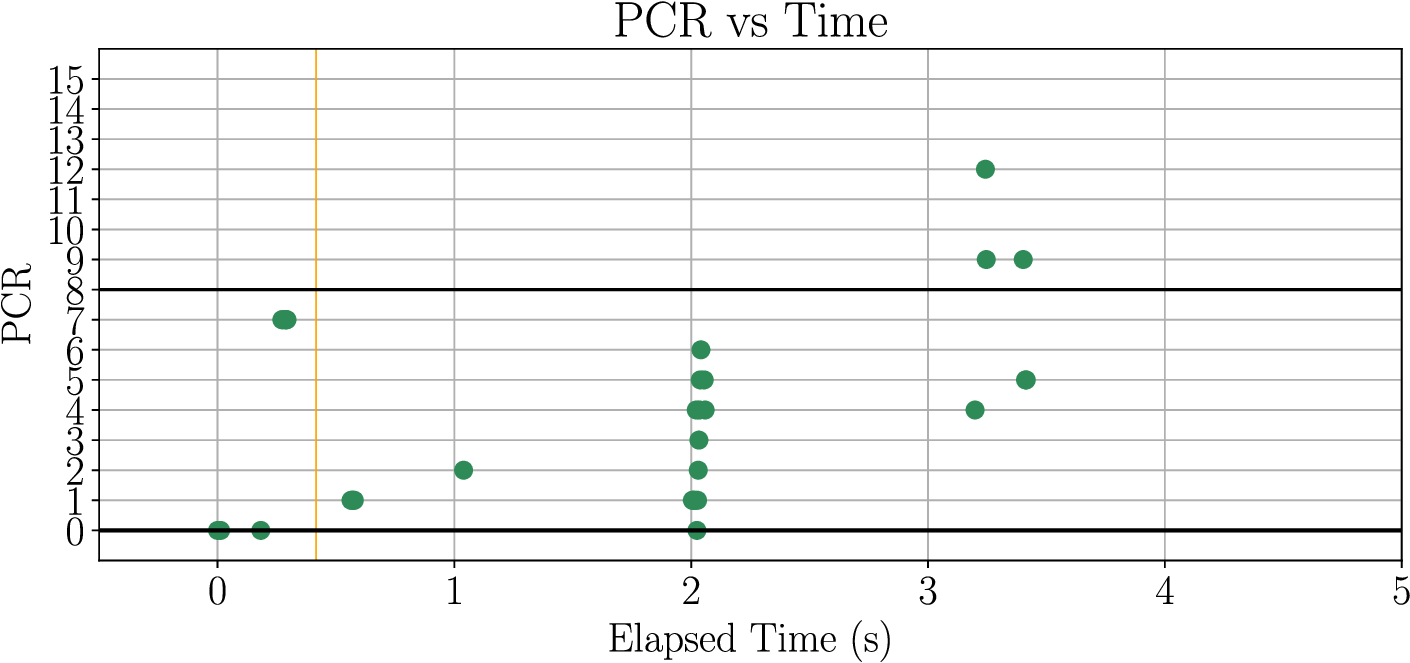}
			\put(66, 31) {
				\begin{tikzpicture}[overlay]
					\draw[pointer] (0.5, 0.25) -- (0.25, 0.0);
				\end{tikzpicture}
			}
		\end{overpic}
		\caption{NixOS, \sd{} v253}
		\label{fig:sd253}
	\end{subfigure}
	\begin{subfigure}[t]{.49\textwidth}
		\centering
		\begin{overpic}[width=1\linewidth,grid=false]{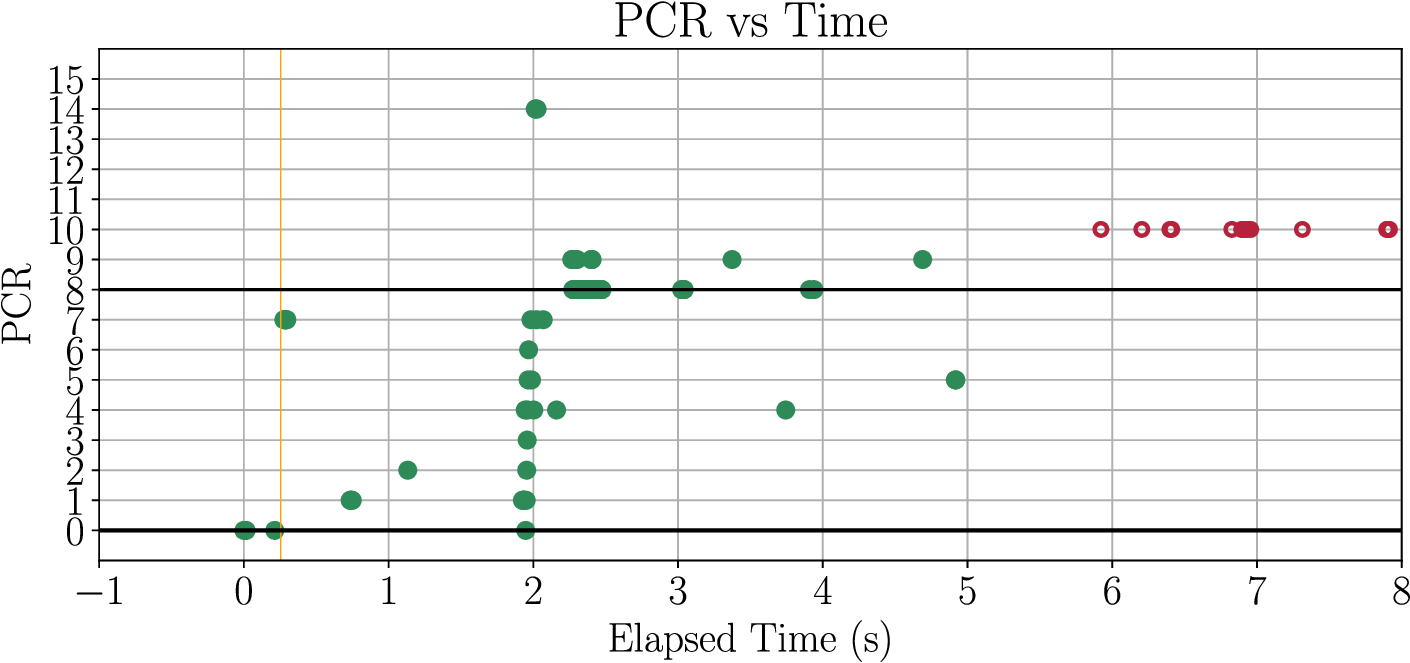}
			\put(35, 33) {
				\begin{tikzpicture}[overlay]
					\draw[pointer,DodgerBlue] (0.5, 0.1) -- (0.25, 0.3);
				\end{tikzpicture}
			}
			\put(66, 31) {
				\begin{tikzpicture}[overlay]
					\draw[pointer] (0.25, 0.3) -- (0.5, 0.1);
				\end{tikzpicture}
			}
		\end{overpic}
		\caption{Fedora 43, \sd{} v258}
		\label{fig:f43sd258}
	\end{subfigure}
	\caption{
        PCR Extend operations captured by TPMSpy.
		Nodes denote PCR Extend commands in the TPMSpy capture; those with a corresponding TPM Event Log entry are green.
        In v245, the PCR 8 measurement (blue arrow) is undocumented behaviour expected since v248.
        In v251, \sd[boot] moved to use PCR 12 (red arrow) instead of PCR 8 (red circle); Linux measurements in PCR 9 (the same red circle) are missing.
        In v253, Linux Kernel extends PCR 9 (red arrow).
        Fedora 43 features Shim by default in PCR 14 (blue arrow), GRUB in PCR 8, and IMA, which extends PCR 10 multiple times (red arrow).
    }
	\label{fig:sdbasic}
\end{figure}

With the first setup, we captured measurements of \sd{} from v245 to v258 and compared them to the documentation analysed in \Cref{subsec:doc}.
\Cref{fig:sdbasic} shows PCR Extend operations of \sd{} versions where significant changes in behaviour were observed.

TPMSpy captures every measurement, including PCRs 0--7 used by low-level firmware components.
However, they lack the detailed documentation required for the same level of analysis we performed for \sd{}; we only compare the measurements between runs to detect unexpected changes.
The graphs show a discrepancy with the documented behaviour: v245 extends PCR 8 beyond the documented behaviour, which first appears in v248.
The graph for v249 contains the same measurements and is omitted for brevity.
According to the commit history in \sd{}'s GitHub, support for measuring the kernel command line was merged into \sd[boot] in the commit \texttt{92ed3bb4}, dated February 2016, and was released with \sd{} v230 unannounced~\cite{systemd:v230-released}.
It remained undocumented until v248, in which TPM is first mentioned in the changelog~\cite{systemd:v248-released}.
Still, if remote attestation were performed for these versions of \sd[boot], the appraisers would need to investigate the source of this measurement.

In \Cref{fig:sd251}, \sd[boot] switched from PCR 8 to PCR 12.
The documentation for v251 also notes that, since version 5.17, the Linux Kernel has extended PCR 9; however, no such measurements are visible in this figure.
This is because NixOS used Linux Kernel 5.15 until February 2023, when it switched to 6.1, and later that month included \sd{} v253.

To analyse features unavailable in NixOS, we captured TPM Fedora 41--43 (v256--v258) (v258 shown in \Cref{fig:f43sd258}) and Ubuntu 24 LTS (v255 in \Cref{fig:u24}).
Instead of \sd[boot], these systems use Shim to verify and start GRUB, and thus we observe extensions to PCR 14 and PCR 8, as predicted by the analysis.
Kernel images are not yet unified in these versions, so \sd{}'s user-space tools do not extend any registers.
On the other hand, we observe IMA measurements in PCR 10, which are not present in the TPM Event Log.
IMA can produce hundreds of measurements, which may exceed the limited memory reserved for the platform TPM Event Log, and are accessible as a virtual file in the kernel's \texttt{sysfs}.
Appraisers wishing to validate PCR 10 must be aware of this and request the file along with the TPM Quote, or derive a set of acceptable values beforehand.

To demonstrate the independence of the TPMSpy tool from the platform, we also captured Windows 11 with BitLocker (\Cref{fig:win11}), where no equivalent documentation exists but a distinct pattern using PCR 11--14 is visible.

We repeated experiment 1,000 times for each \sd{} version in NixOS.
We compared the traces and metadata to verify the deterministic order of events, the absence of race conditions, and the absence of unexpected changes in measured values.
To measure impact on system performance, we observed 10 runs of Fedora 43 (which has more features enabled than NixOS) from startup to \sd{} reporting readiness, taking $22.13 \pm 1.26 \text{s}$.
Without TPMSpy, the same setup took $23.77 \pm 1.48 \text{s}$; the $1.64 \text{s}$ is comparable to variability and thus not statistically significant.
Direct resource measurements showed TPMSpy using only $40.073 \pm 3.387 \text{ms}$ of CPU time, confirming negligible overhead.

\subsection{Setup 2: Evaluation of Setup with Unified Kernel Image}
\label{subsec:setup_uki}

In the second setup, we installed Unified Kernel Image and enabled LUKS volume key measurement.
In the case of Fedora, we also replaced Shim and Grub with \sd[boot] to enable user-space measurements.

\begin{figure}
	\centering
	\begin{subfigure}[t]{.49\textwidth}
		\centering
		\begin{overpic}[width=1\linewidth,grid=false]{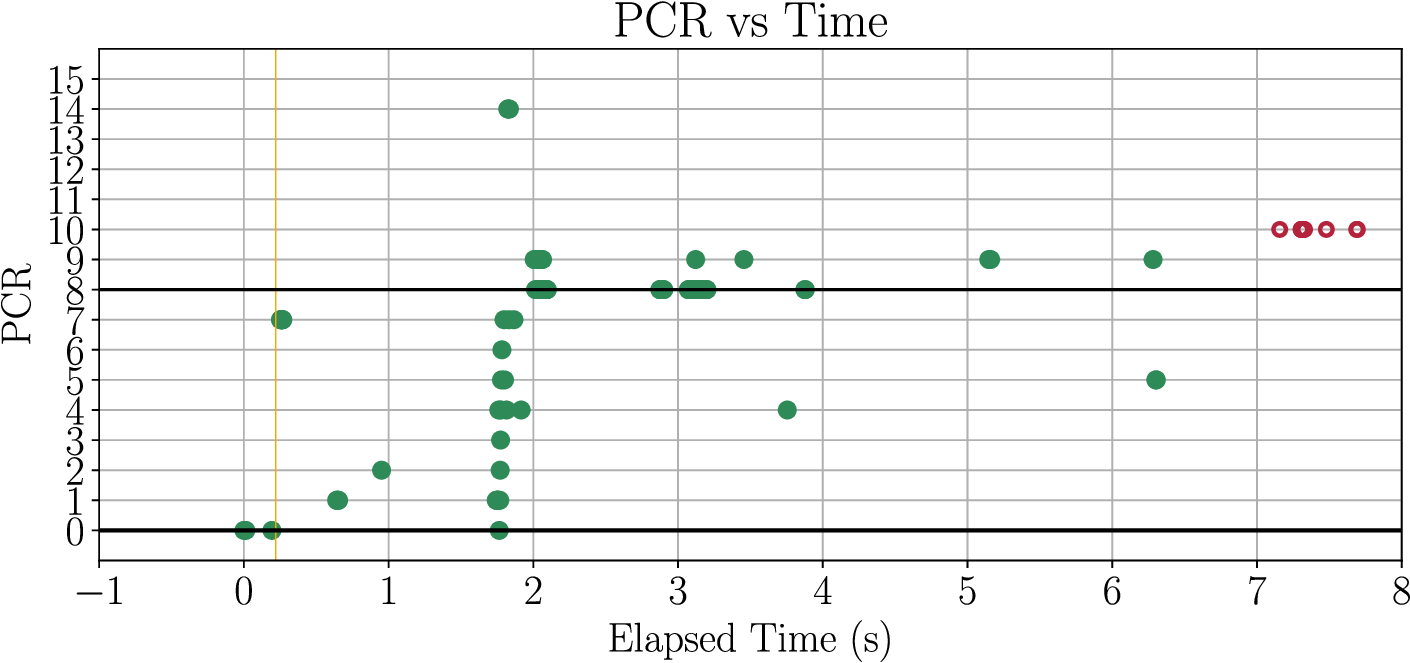}
		\end{overpic}
		\caption{Ubuntu 24, \sd{} v255}
		\label{fig:u24}
	\end{subfigure}
	\begin{subfigure}[t]{.49\textwidth}
		\centering
		\begin{overpic}[width=1\linewidth,grid=false]{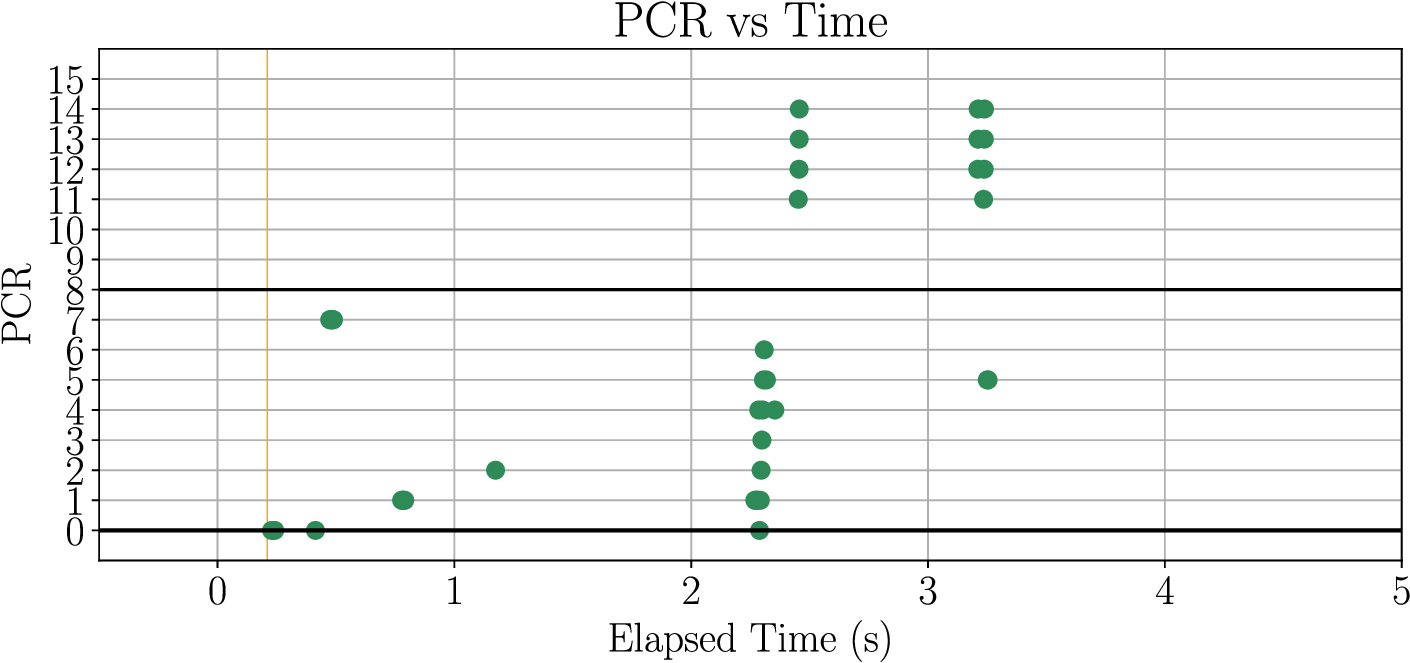}
			\put(44, 37) {
				\begin{tikzpicture}[overlay]
					\draw[pointer] (0.25, 0.3) -- (0.5, 0.1);
                    \draw[pointer] (1.05, 0.3) -- (1.30, 0.1);
				\end{tikzpicture}
			}
		\end{overpic}
		\caption{Windows 11, no \sd{}}
		\label{fig:win11}
	\end{subfigure}
    \par\bigskip
	\begin{subfigure}[t]{.49\textwidth}
		\centering
        \begin{overpic}[width=1\linewidth,grid=false]{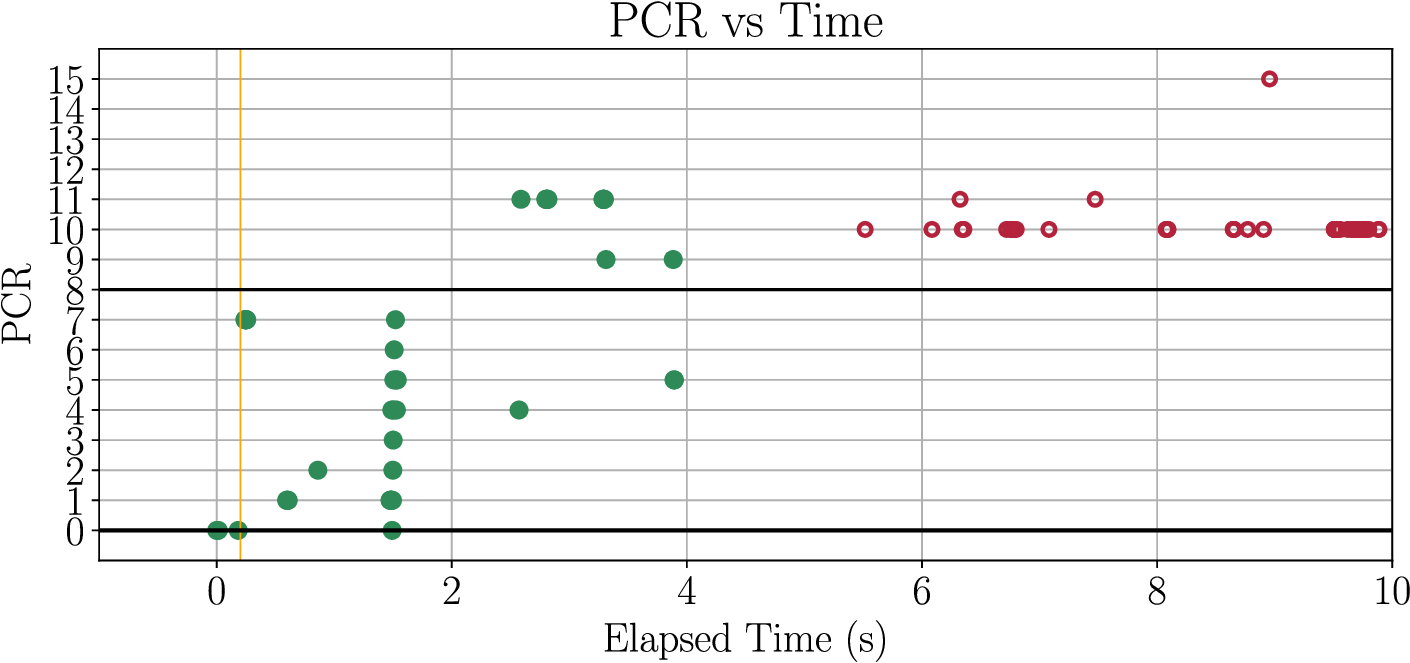}
			\put(75, 36) {
				\begin{tikzpicture}[overlay]
					\draw[attention] (-0.1, 0.30) circle;
					\draw[pointer,DodgerBlue] (0.45, 0.0) -- (0.70, 0.25);
					\draw[pointer,DodgerBlue] (-0.85, 0.15) -- (-0.60, -0.10);
				\end{tikzpicture}
			}
		\end{overpic}
		\caption{Fedora 43, v258, UKI}
		\label{fig:f43uki}
	\end{subfigure}
	\begin{subfigure}[t]{.49\textwidth}
		\centering

		\begin{overpic}[width=1\linewidth,grid=false]{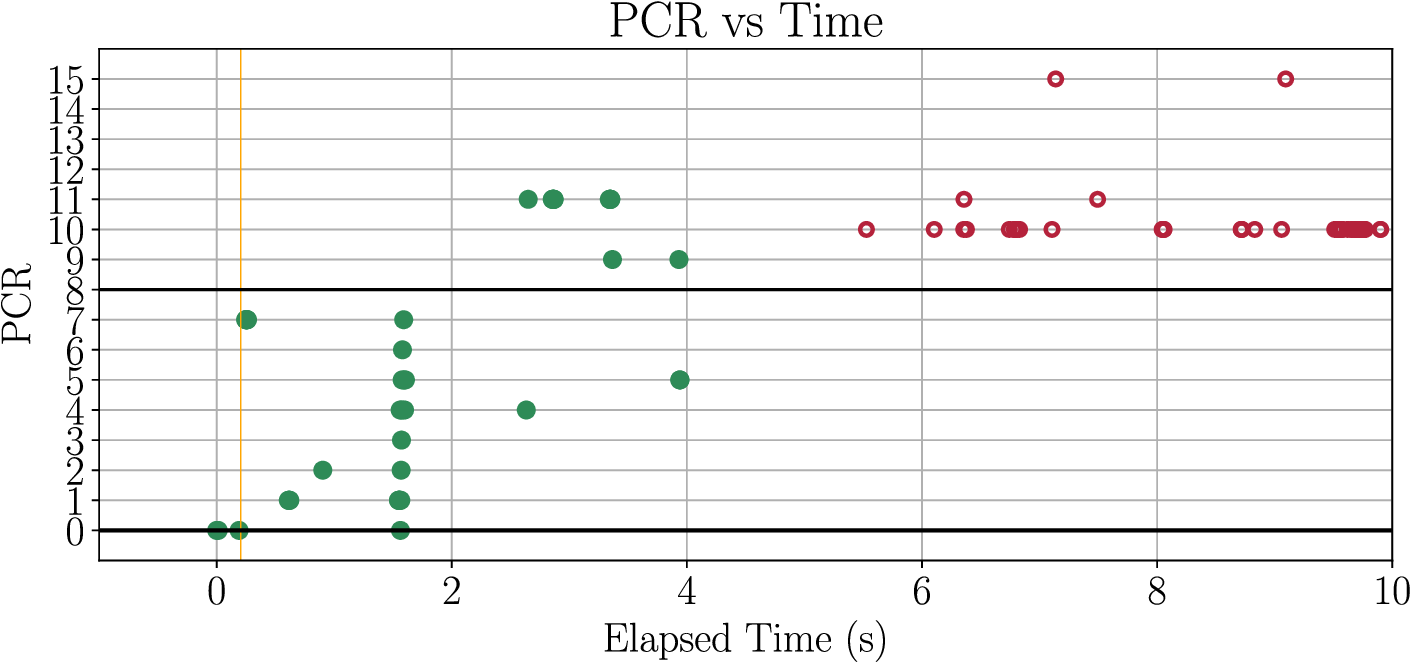}
			\put(63, 36) {
				\begin{tikzpicture}[overlay]
					\draw[pointer] (0.25, 0.0) -- (0.5, 0.25);
					\draw[pointer,DodgerBlue] (1.25, 0.0) -- (1.5, 0.25);
					\draw[pointer,DodgerBlue] (-0.15, 0.15) -- (0.10, -0.10);
				\end{tikzpicture}
			}
		\end{overpic}

		\caption{LUKS volume key measurement}
		\label{fig:f43uki15}
	\end{subfigure}

	\caption{
        Top: PCR Extend operations of \sd{} v255 in Ubuntu 24 (left) and Windows 11 (right).
        The documentation for Windows' measurements is not specific enough to allow PCR analysis we performed for \sd{}, but it shows a distinct pattern (red arrows) not seen with \sd{}.\\
        Bottom: PCR extends on \sd{} v258 with Unified Kernel Image.
        Aside from IMA measurements, user-space utilities extend PCR 11 and PCR 15 (blue arrows).
        An earlier measurement to PCR 15 is missing in the default setting (left, red circle) but appears when enabled explicitly (right, red arrow).
    }
\end{figure}

\Cref{fig:f43uki} shows Fedora 43 with v258, UKI and default kernel parameters.
According to the documentation and \Cref{fig:sdusage}, PCRs 9, 11, 12, and 15 should be extended.
Early measurements to PCR 11 (green nodes) accounted for by TPMSpy and TPM Event Log are issued by \sd[stub], which measures parts of the kernel and the initial file system.
Later PCR 11 measurements are submitted by \sd[pcrphase] and are not recorded in the TPM Event Log.
PCR 12 measurements of the kernel command by \sd[boot] are missing.
There is one PCR 15 measurement by \sd[pcrmachine], though the documentation suggests that \sd[pcrfs] should have submitted measurements as well.
In \Cref{fig:f43uki15}, we enabled the kernel parameter for \sd[cryptsetup] to extend PCR 15 with the encrypted volume key.
This time, the measurement was detected by TPMSpy but not recorded in the TPM Event Log.

The v255 changelog announces that user-space measurements from \sd{} are logged into \texttt{/run/log/systemd/tpm2-measure.log}, and that these measurements were previously logged to the journal only~\cite{systemd:v255-released}.
The TPM Event Log alone is thus insufficient for remote attestation: PCR 11 (v252), PCR 15 (v253--v254) and PCR 10 (IMA) measurements appear in specific files only.

In this case, TPMSpy provides a comprehensive set of observations in a single capture, enabling the system developer to trace all event logs stored in the system and document them for appraisers.

\section{Conclusion}
\label{sec:conclusion}

In this paper, we propose a novel methodology for the independent verification of the TPM Event Log based on virtualization.
We designed and implemented TPMSpy, a virtual TPM interposer, which observes the complete communication between a virtual system and a software TPM emulator.
Other available methods either require specialized hardware, focus on specific parts of the boot process, or verify shorter boot chains.
In contrast, our method enables repeated, automated, and non-intrusive assessment for every stage of the boot process, independent of the system, as demonstrated by capturing runs of NixOS, Fedora, Ubuntu and Windows.
It thus allows for the systematic evaluation of Measured Boot on any platform that features a TPM, including those without available source code.

We observed an increasing use of PCR over time, which agreed with the documentation in most cases.
We also observed deviations from expected behaviour for \sd{} user-space services and IMA, which store their measurement metadata outside the TPM Event Log.
This deviation affects remote attestation: an appraiser unaware of these locations will fail to attest to the measured values.
Our method enables the independent observation of TPM commands and responses, providing complete coverage of TPM interactions during boot.

The strict independence from the observed system limits available context: currently, only timestamps and channel metadata are recorded along with the captured events.
Future research may address this limitation by querying the virtual machine's state at the time of event capture and, ideally, precisely identifying the component that originated the event.
Furthermore, the method verifies TPM Event Log completeness but does not replace formal verification, as it cannot guarantee that every component executed during boot was measured.
The absence of a specific measurement will not cause a discrepancy between the TPM Quote and the TPM Event Log, and thus cannot be detected solely by our method.
Future research can address this limitation by deriving expectations from static code analysis or by verifying that every piece of code loaded into memory is measured, as wide divergence warrants systematic analysis.

\begin{credits}
	\subsubsection{\ackname} The authors of this paper were supported by the EU project CHESS \#101087529.

	\subsubsection{\discintname}
The authors have no competing interests to declare that are relevant to the content of this article.
\end{credits}

\clearpage
\bibliography{tpmspy}

\end{document}